text of paper:

\documentstyle[preprint,aps]{revtex}
\begin{document}
\preprint{ETH-TH/ 93-50}
\title{
Luttinger Liquid in a Solvable 2-Dimensional Model.$^{*}$
}
\author{R. Hlubina}
\address{Theoretische Physik, ETH-H\"onggerberg,
CH-8093 Z\"urich, Switzerland}
\maketitle
\begin{abstract}
\noindent
We consider spinless electrons in two dimensions with the bare spectrum
$\epsilon({\bf p})=|p_x|+|p_y|$. In momentum space, the interactions among
electrons have a finite range $q_0$, which is small compared to the Fermi
momentum. A golden rule calculation of the electron lifetime indicates a
breakdown of the Landau Fermi liquid in the model. At the one-loop level of
perturbation theory, we show that the density wave and the superconducting
instabilities cancel each other and there is no symmetry breaking. We solve
the model via bosonization; the excitation spectrum is found to consist of
gapless bosonic modes as in a one-dimensional Luttinger liquid.
\end{abstract}
\pacs{PACS numbers:}

%
%
%
%INTRODUCTION
%
%
%

Recently, the question about validity of Landau's Fermi liquid (LFL) theory
in two dimensions has attracted much interest. This is because there are
strong indications that in the normal state of the high temperature
superconductors (HTSC's), some kind of a non-LFL state is realized.  This
idea is due to Anderson \cite{Anderson}, who realized that there are
striking analogies between the phenomenology of the cuprates and of
one-dimensional metals. The latter are known to be Fermi liquids of a very
special type (Luttinger liquids): the Fermi surface (defined as a set of
points, where the momentum distribution function has a singularity) still
exists and has the same volume as for non-interacting electrons in agreement
with Luttinger's theorem, but elementary excitations are completely
different from those of a usual metal in $D=3$: there are no quasiparticle
excitations and the spectrum consists of collective modes only. However, in
a $D=2$ low density Fermi gas with short ranged interactions, the LFL phase
was shown (within perturbation theory) to be stable \cite{Engelbrecht}. It
is believed, that LFL-behavior can break down in perturbation theory only in
specific situations: assuming particular bandstructure (like in the nested
Fermi liquid theory \cite{Ruvalds}) and/or singular long ranged
interactions. The latter suggestion is due to Haldane \cite{Haldane} (see
also \cite{HM}), who proposed a generalization of bosonization as a tool to
study non-LFL behavior in higher dimensions.  Haldane's method was applied
recently to study fermions interacting with long-ranged current-current
interactions
\cite{KHR}.

%
%
%
%MODEL
%
%
%

Here we study a model with nesting. We consider spinless fermions in $D=2$
described by the Hamiltonian
\begin{equation}
H=\sum_{{\bf p}} \epsilon_{\bf p} c^\dagger_{{\bf p}}c_{{\bf p}}+
{1\over{2\Omega}}\sum_{{\bf p},{\bf p}^\prime,{\bf q}} V({\bf q})
c^\dagger_{{\bf p}+{\bf q}} c^\dagger_{{\bf p}^\prime-{\bf q}} c_{{\bf
p}^\prime}c_{{\bf p}},
\end{equation}
where $\epsilon_{\bf p}=v_F\:(|p_x|+|p_y|)$ is the bare electron spectrum,
$V({\bf q})$ is a Fourier transform of the electron-electron interaction and
$\Omega$ is the (two-dimensional) volume of the system \cite{toy}. There are
4 branches $\alpha=1,..,4$ of the spectrum, one in each quadrant of the
${\bf p}$ plane (see Fig.1).  The group velocity in branch $\alpha$ is ${\bf
v}_\alpha=\partial\epsilon^\alpha_{\bf p}/\partial{\bf p}$.  Our dispersion
corresponds to fermions moving in a planar quadratic lattice described by
the hopping Hamiltonian $H_{kin}=\sum_{{\bf i},{\bf R}}t_{{\bf
R}}c^\dagger_{\bf i}c_{\bf i+R}$, where $t_{\bf R}/v_F=\pi\delta_{m,0}
\delta_{n,0} +\delta_{n,0}((-1)^m-1)/\pi m^2+\delta_{m,0}((-1)^n-1)/
\pi n^2$.  We denoted ${\bf R}=(m,n)$;
$\delta_{m,n}$ is the Kronecker symbol. It is seen that the chosen band
structure is quite unrealistic, since it requires that the overlaps of
localized orbitals fall off only quadratically with distance.  The
motivation for studying the model Eq.(1) is twofold: the linearity of the
spectrum allows for a solution of the model via bosonization; on the other
hand, the Fermi surface of the model Eq.(1) exhibits the property of perfect
nesting without being complicated by the occurence of van Hove
singularities, as is the case for the nearest neighbor tight binding model.
Thus it provides a convenient testing ground for nested Fermi liquid theory.

Let us assume now that $V({\bf q})=V\Theta(q_0-|{\bf q}|)$, $\Theta(x)$ is
the Heaviside function. We take $q_0\ll\Lambda$, where $\Lambda$ is the
length of the Fermi surface in one of the branches (see Fig. 1). For what
follows, it is sufficient to consider potentials with finite range in
momentum space and our choice is dictated by convenience only.  In real
space, our potential falls off at large distances $q_0r\gg1$ as
$V(r)=(2\sqrt{2/\pi}) U\sin(q_0r-\pi/4)/(q_0r)^{3/2}$, where $U=Vq_0^2/4\pi$
stands for the contact interaction energy.

%
%
%
%LIFETIME
%
%
%

We calculate the lifetime of an electron using Fermi's golden rule.  There
are three types of scattering possible: an electron ${\bf p}$ scattering to
the state ${\bf p+q}$ in branch 1 can excite an electron-hole pair ${\bf k}$
and ${\bf k-q}$ in branch 3, branch 1 or branch 2 (4). We denote the
corresponding processes by $g_2$, $g_4$ and $g_{FL}$.  The $g_{FL}$
processes do not make use of the peculiarities of the bandstructure and lead
to the scattering rate $1/\tau\approx G^2\epsilon^2/E_F$ of Fermi liquid
type, where $G=(1/2\pi)^2\Lambda V/v_F$ is a dimensionless coupling
constant.  For the $g_2$ processes, the scattering rate in the limit
$\epsilon,\epsilon_{\bf p}\ll v_Fq_0$ becomes
\begin{eqnarray}
{1\over{\tau_{\bf p}(\epsilon)}}=\pi\:
{{q_0}\over{\Lambda}}\:G^2\:
(\epsilon-\epsilon_{\bf p})\Theta(\epsilon-\epsilon_{\bf p}),
\end{eqnarray}
demonstrating the well-known linear dependence of the lifetime on energy.
Consider now the $g_4$ processes. Since the Fermi surface is a line and not
a point, scatterings in this channel do not cancel as in the case of
spinless fermions in $D=1$, and there will be only partial cancellation of
such scatterings due to Pauli principle.  This happens if all momenta
involved in the scattering process differ by $q_0$. In our case
$q_0\ll\Lambda$ and cancellations are completely irrelevant.  For
$\epsilon,\epsilon_{\bf p}\ll v_Fq_0$, we find
\begin{eqnarray}
{1\over{\tau_{\bf p}(\epsilon)}}=2\pi\: {{q_0}\over{\Lambda}}\:G^2\:
\epsilon_{\bf p}^2\delta(\epsilon-\epsilon_{\bf p}),
\end{eqnarray}
where $\delta(x)$ is a Dirac delta-function. Eq.(3) is analogous to the
scattering rate of electrons with spin in $D=1$ in the $g_4$ channel
\cite{Metzner}. (In one dimension, this golden rule result indicates the
spin-charge separation found in the exact solution \cite{Meden}. In fact, by
Kramers-Kronig analysis one finds that the real part of the self-energy is
$\Re\Sigma({\bf p},\epsilon)\sim 1/(\epsilon-\epsilon_{\bf p})$. The
`quasiparticle' spectrum is given by the solutions of
$\epsilon-\epsilon_{\bf p}-\Sigma({\bf p},\epsilon)=0$. There are two of
them, corresponding to spin and charge excitations.)  In the present case,
the spin degree of freedom is replaced by the position within the branch
and, as a result of $g_4$-like scatterings, coherent propagation of
electrons is again lost.  A similar golden rule calculation was performed in
the context of nested Fermi liquid theory of HTSC's \cite{Ruvalds}.
Unfortunately, the authors of Ref.\cite{Ruvalds} missed the contribution of
$g_4$-like processes to the scattering rate and, as a consequence, also the
spin-charge separation.

%
%
%
%INSTABILITIES
%
%
%

Due to the perfect nesting of the Fermi surface, the model Eq.(1) shows a
strong tendency towards the charge density wave (CDW) formation. This state
is characterized by a nonzero expectation value of the operator
$D^\dagger_{\bf p}=c^\dagger_{{\bf p}+{\bf Q}}c_{\bf p}$, describing a
particle-hole pair with total momentum ${\bf Q}$ (where ${\bf Q}$ is the
nesting vector of the Fermi surface, see Fig.1). It is important to note
that there are in general two nesting vectors ${\bf Q}_1$ and ${\bf Q}_2$ of
the Fermi surface. We restrict ourselves to the case of a half filled band,
for the moment. In that case ${\bf Q}_2={\bf Q}_1+{\bf G}$ where ${\bf G}$
is an inverse lattice vector and ${\bf Q}_2\equiv{\bf Q}_1\equiv{\bf Q}$.
This in turn means that an electron-hole pair $D^\dagger_{\bf p}$ can be
scattered around the whole Fermi surface. Rewriting the interaction term of
the Hamiltonian, the scattering in the CDW channel is found to be $-V({\bf
p},{\bf p^\prime})\:D^\dagger_{\bf p^\prime}\:D_{\bf p}$.  Assuming that a
charge density wave is formed, we can write $\langle D^\dagger_{\bf
p}\rangle=d_{\bf p}$ (where $\langle...\rangle$ denotes the thermal
expectation value) and the mean field equation for the order parameter
$d_{\bf p}$ reads
\begin{equation}
d_{\bf p}={1\over\Omega}\sum_{\bf k}\:V({\bf p},{\bf k})\: {{d_{\bf
k}}\over{2E_{\bf k}}}\:\tanh{{E_{\bf k}}\over{2T}},
\end{equation}
where $E_{\bf p}=\sqrt{\epsilon_{\bf p}^2+d_{\bf p}^2}$.  Repeating the
above analysis for the superconducting instability, we assume pairs of
particles $P^\dagger_{\bf p}=c^\dagger_{\bf p} c^\dagger_{-\bf p}$ to be
scattered coherently around the Fermi surface.  The relevant matrix element
is $V({\bf p},{\bf p^\prime})\:P^\dagger_{\bf p^\prime}\:P_{\bf p}$ and the
equation for the expectation value of the order parameter $\langle
P^\dagger_{\bf p}\rangle=\Delta_{\bf p}$ has the familiar BCS form
\begin{equation}
\Delta_{\bf p}=-{1\over\Omega}\sum_{\bf k}\:V({\bf p},{\bf k})\:
{{\Delta_{\bf k}}\over{2E_{\bf k}}}\:\tanh{{E_{\bf k}}\over{2T}}
\end{equation}
with $E_{\bf p}=\sqrt{\epsilon_{\bf p}^2+\Delta_{\bf p}^2}$. Note that the
sign in front of the interaction term $V({\bf p},{\bf k})$ is different for
the CDW and BCS instabilities. To determine the kind of instability which
occurs (on the mean field level), one has to find all nontrivial solutions
to Eqs.(4,5) and determine the corresponding $T_c$ values. The instability
with the highest value of $T_c$ is dominant \cite{dwave}.

In order to decide whether the instabilities are real, we investigate two
body scattering vertices in the corresponding channels and investigate their
scaling in one loop approximation. The mean field approximation corresponds
to restricting ourselves to one particular one loop diagram.  We will show
that taking into account {\it all} diagrams of the given order completely
changes the result and instead of a logarithmic flow, one observes that the
scattering vertex stays constant suggesting a line of fixed points in our
toy model, as is the case for $D=1$ \cite{Solyom}.  Let us start with the
CDW channel. There are 5 one loop diagrams for scattering from
$D^\dagger_{\bf p}$ to $D^\dagger_{\bf p^\prime}$: (a) the ladder diagram,
(b) the crossed diagram, (c)+(d) vertex corrections and (e) the screening
diagram.  We set ${\bf p}$ and ${\bf p^\prime}$ to lie on the Fermi surface
and the energies of the incoming and outgoing electrons are 0.  We work at
finite temperatures.  The diagram (a) is the one considered in the mean
field theory. Its contribution is precisely cancelled by (b). In fact, the
condition $q_0\ll\Lambda$ implies small momentum transfer scatterings and
that is why ${\bf p}\approx{\bf p^\prime}\approx{\bf k}$.  For such momenta,
an exact equality $\epsilon_{{-\bf k}+{\bf p}+{\bf p^\prime}}=
-\epsilon_{\bf k}$ holds and, consequently,
\begin{equation}
\Gamma^{(a)}_{{\bf p},{\bf p^\prime}}=
-\Gamma^{(b)}_{{\bf p},{\bf p^\prime}}=\sum_{\bf k}\:
V_{{\bf p},{\bf k}}\:V_{{\bf k},{\bf p^\prime}}\:
{{f(\epsilon_{\bf k})-f(-\epsilon_{\bf k})}
\over{2\epsilon_{\bf k}}}.
\end{equation}
The remaining diagrams (c-e) contribute
\begin{eqnarray}
\Gamma^{(c+d+e)}_{{\bf p},{\bf p^\prime}}(\omega)=
V_{{\bf p},{\bf p^\prime}}\sum_{\bf k} W_{{\bf p},{\bf p^\prime},{\bf k}}
{{f(\epsilon_{\bf k})-f(\epsilon_{\bf k+p^\prime-p})}
\over{\epsilon_{\bf k}-\epsilon_{\bf k+p^\prime-p}-\omega}},
\end{eqnarray}
where $W_{{\bf p},{\bf p^\prime},{\bf k}}=-V_{{\bf p},{\bf p^\prime}}
+V_{{\bf p+q},{\bf k}}+V_{{\bf p},{\bf k}}$ and we have introduced an
infinitesimally small energy transfer $\omega$ into the scattering process.
Evaluating $\Gamma^{(c+d+e)}_{{\bf p},{\bf p^\prime}}(\omega)$ for finite
$\omega$ and taking the limit $\omega\rightarrow0$ gives
$\Gamma^{(c+d+e)}_{{\bf p},{\bf p^\prime}}=0$. Thus the scattering vertex in
the CDW channel does not flow in the one loop approximation.  The analysis
of the BCS channel is analogous.  Summarizing, we do not find any sign of
symmetry breaking neither in the CDW, nor in the BCS channel.

Note that the above argument about the cancellation of the CDW and BCS
instabilities makes use only of the linearity of the spectrum in a given
branch and it is valid for any form of the interaction, as long as the
around the corner processes can be neglected; moreover, it applies also away
from half filling.  The uncontrolled point of our analysis is the neglect of
`around the corner' processes, which should be studied more carefully.

We expect that taking the absence of the around the corner processes for
granted, our results will stay true in all orders of perturbation theory,
since this neglect results in an additional conservation law: the number of
electrons in a given branch is unchanged in the scattering. In $D=1$,
Dzyaloshinskii and Larkin \cite{LarkinDzya} used conservation of electrons
in a given branch and the corresponding Ward identities to solve the
Tomonaga-Luttinger model exactly.  Recently, the importance of conservation
laws in investigating the non-LFL behavior in higher dimensions was stressed
in several studies \cite{Haldane,HM,MetznerCastro}.

%
%
%
%BOSONIZATION
%
%
%

Finally, we solve the model via Haldane's bosonization \cite{Haldane}.  In
bosonizing the kinetic energy $H_{kin}=\sum_{\bf p} \epsilon_{\bf p}
{c^\dagger}_{\bf p} c_{\bf p},$ we closely follow the standard methods in
$D=1$ \cite{MattisLieb}.  For every branch $\alpha$, we construct a density
operator $\rho_\alpha({\bf q})=\sum_{\bf p}\Theta_\alpha({\bf
p+q})\Theta_\alpha({\bf p})c^\dagger_{{\bf p}+{\bf q}}\:c_{\bf p}$, where
$\Theta_\alpha({\bf p})=1$, if ${\bf p}$ lies in the quadrant $\alpha$, and
$\Theta_\alpha({\bf p})=0$ otherwise. Note that $D=2$ is different from the
one-dimensional case in that even restricting ourselves to the neighborhood
of the Fermi surface, pairs ${\bf p}$ and ${\bf p}+{\bf q}$ lying in
different branches exist for small ${\bf q}$.  This is due to the fact that
the Fermi surface in $D>1$ is a continuous manifold. If we want to interpret
the physical density operator $\rho({\bf q})$ as $\rho({\bf
q})=\sum_\alpha\rho_\alpha({\bf q})$, we have to neglect the excitations of
the electrons between different branches. This is a justified approximation,
if the number of neglected processes is much less than their total number,
i.e. we have to restrict ourselves to the study of sufficiently long
wavelength processes, $|{\bf q}|\:\ll\:\Lambda$.  Neglecting the above
mentioned `around the corner' processes and using the linearity of the
spectrum $\epsilon_{\bf p}$, one has
\begin{equation}
[H_{kin},\rho_\alpha({\bf q})]= {\bf v_\alpha}\cdot{\bf q}\:
\rho_\alpha({\bf q}),
\end{equation}
straightforwardly generalizing the result of the one-dimensional
calculation. Closely following the derivation of the commutator of the
density operators in $D=1$ (see, e.g., Ref.\cite{Haldane81}), we find
$[\rho_\alpha({\bf q}),\rho_\beta(-{\bf q}^\prime)]=
\delta_ {\alpha,\beta}\:\delta_{{\bf q},{\bf q}^\prime}\:\sum_{\bf
p}(n_{\bf p+q}^o-n_{\bf p}^o)$, where $n_{\bf p}^o$ is the distribution of
noninteracting fermions. From here we have
\begin{equation}
[\rho_\alpha({\bf
q}),\rho_\beta(-{\bf q}^\prime)]= -\delta_{\alpha,\beta}\:
\delta_{{\bf q},{\bf q}^\prime}\:{{\Lambda\Omega}\over{(2\pi)^2}}
\:{\bf n_\alpha}\cdot{\bf q},
\end{equation}
where ${\bf n_\alpha}$ is a unit vector in the direction of the group
velocity ${\bf v_\alpha}$. Thus the kinetic energy operator can be written
$H_{kin}={{(2\pi)^2v_F}\over{2\Omega\:\Lambda}}\:\sum_{\alpha}
\sum_{\bf q}:\rho_\alpha({\bf q})\rho_\alpha(-{\bf q}):$, where
$:X:$ denotes normal ordering of $X$.  Let us investigate the interaction
energy, which is given by the second term in Eq.(1).  Neglecting the around
the corner processes, we can write
$H_{int}={1\over{2\Omega}}\:\sum_{\alpha,\beta}\:\sum_{\bf q}\: V({\bf
q})\::\rho_\alpha({\bf q})\:\rho_\beta(-{\bf q}):$. Introducing new
operators via $\rho_\alpha({\bf q})= a_\alpha({\bf
q})\sqrt{\Omega\Lambda/(2\pi)^2}$, the total Hamiltonian can be written
\begin{eqnarray}
H={1\over 2}\sum_{\bf q}v_F|{\bf q}|\sum_{\alpha, \beta}\:
\left(\delta_{\alpha,\beta}+G\right)\:\sqrt{|\cos\alpha\:\cos\beta|}\:
:a_\alpha({\bf q})\:a_\beta(-{\bf q}):.
\end{eqnarray}
The commutation relations for $a_\alpha({\bf q})$ are $[a_\alpha({\bf
q}),a_\beta(-{\bf q}^\prime)]= -\delta_{\alpha,\beta}\:\delta_{{\bf q},{\bf
q}^\prime}$sign$\left ({\bf n_\alpha}\cdot {\bf q}\right)$.  Thus
$a_\alpha({\bf q})$ is a Bose creation operator for sign$\left({\bf
n_\alpha}\cdot {\bf q}\right)=1$ and an annihilation operator otherwise.  We
used $\alpha$, resp. $\beta$ to denote the angles between ${\bf q}$ and
${\bf n}_\alpha$, resp. ${\bf n}_\beta$.  To proceed further, we have to
choose the direction of ${\bf q}$ in order to distinguish the creation and
annihilation operators. It is easy to see that all ${\bf q}$, whose angle
with the $p_y$ axis lies in the interval $(-\pi/4,\pi/4)$, lead to the same
choice of operators. Ignoring the ground state energy and concentrating on
the excitations, we have after symmetrization $H'={1\over 4}\sum_{\bf
q}\:v_f|{\bf q}|\:A^\dagger({\bf q}){\bf M}({\bf q}) A({\bf q})$, where $H'$
is the part of the Hamiltonian corresponding to the chosen range of ${\bf
q}$, $A^\dagger({\bf q})=
(a_1^\dagger,a_3^\dagger,a_2^\dagger,a_4^\dagger,a_1,a_3,a_2,a_4)$ and ${\bf
M}$ is a real symmetric $8\times 8$ matrix, $${\bf
M}=\left(\begin{array}{ll} {\bf M}_1&{\bf M}_2\\ {\bf M}_2&{\bf M}_1
\end{array}\right).$$
The matrices ${\bf M}_1$ and ${\bf M}_2$ are given in terms of
$A=(1+G)\cos\phi$, $B=(1+G)\sin\phi$, $C=G\cos\phi$,
$D=G\sin\phi$ and $E=G\sqrt{\sin\phi\cos\phi}$ (where $\phi$ is
the angle between ${\bf n}_1$ and ${\bf q}$) by
\begin{eqnarray*}\begin{array}{cc}
{{\bf M}_1=\left(\begin{array}{cc}
A&E\\
E&B
\end{array}\right)\otimes{\bf 1}}&
{\hspace{1cm}{\bf M}_2=\left(\begin{array}{cc}
C&E\\
E&D
\end{array}\right)\otimes\sigma^2,}
\end{array}
\end{eqnarray*}
where $\sigma^2$ is a Pauli matrix and $\otimes$ stands for a direct
product. The Hamiltonian Eq.(10) is quadratic in Bose operators and can be
diagonalized by a generalized Bogoliubov transformation.  This problem can
be reduced to a standard eigenvalue problem for the matrix ${\bf N}=({\bf
M}_1-{\bf M}_2)\:({\bf M}_1+{\bf M}_2)$ \cite{Hiro} and the Hamiltonian
Eq.(10) can be written
\begin{equation}
H=\sum_{\bf q}\:v_f|{\bf q}|\sum_{\alpha=\pm}
\lambda_{\alpha}b_\alpha^\dagger({\bf q}) b_\alpha({\bf q}),
\end{equation}
where $\lambda_{\pm}^2$ are the eigenvalues of ${\bf N}$,
\begin{equation}
\lambda_{\pm}^2={1\over 2}\left[1+2G\pm
\sqrt{(1+4G)\cos^2(2\phi)+(2G)^2}\right].
\end{equation}
Note that $\lambda_{\pm}$ as given by Eq.(12) can be defined in the whole
${\bf q}$ plane by the same formula. That is why we omitted in Eq.(10) the
prime on the Hamiltonian and the sum is over the whole ${\bf q}$ plane.  In
the noninteracting case $G=0$ and $\lambda_{\pm}$ reduce to
$|\sin\phi|$ and $|\cos\phi|$, giving the correct excitation energies ${\bf
v}_\alpha\cdot{\bf q}$ for the collective modes. Eqs.(11,12) represent a
solution to the model Eq.(1). The low energy spectrum is found to consist of
gapless bosonic modes. The quasiparticles are dissolved in these collective
excitations.

%
%
%
%CONCLUSIONS
%
%
%

We would like to point out that it is possible to embed the model  spectrum
studied here into a sequence of spectra $\epsilon_n({\bf k})=(1/2m)
\left(k_x^{2n}+k_y^{2n}\right)^{1/n}$ with $n=1,2,...$ as its limit
$n\rightarrow\infty$. In fact, the constant energy line equation
$\epsilon_\infty({\bf k})=\epsilon$ has the solution
$|k_x|=\sqrt{2m\epsilon}$, $k_y=0$ or $|k_y|=\sqrt{2m\epsilon}$, $k_x=0$ and
the spectrum $\epsilon_\infty({\bf k})$ is (in the neighborhood of the Fermi
energy $\epsilon_F$) identical to that studied here, if one requires
$v_F=\sqrt{2\epsilon_F/m}$.  Decreasing $n$, the Fermi surface becomes more
and more curved, until one finally obtains the usual isotropic spectrum
$\epsilon_1({\bf k})={\bf k}^2/2m$. The crossover between the Luttinger
liquid for $n=\infty$ and Landau's Fermi liquid for $n=1$ will be studied
elsewhere \cite{lifetime}.

We emphasize that the above analysis does not apply to a half filled nearest
neighbor tight binding band. In fact, an important point in deriving the
cancellation of the CDW and BCS instabilities is that the bare spectrum does
not depend (within a given branch) on the direction perpendicular to ${\bf
v}_{\alpha}$, which is not true for a tight binding model. Moreover, in a
nearest neighbor tight binding band, there are van Hove singularities
exactly in the corners of the Fermi surface leading to a completely
different picture than the one used here. In fact, let us consider electrons
within the energy shell $\Delta$ around the Fermi surface. In our model, the
ratio of the number of `corner' states in the shell to the total number of
states within the shell is $\sim\Delta/v_F\Lambda\rightarrow 0$, as
$\Delta\rightarrow 0$. This justifies the neglect of the corner states in
our calculation.  However, in a nearest-neighbor tight binding model, the
fraction of the `corner' states remains finite for $\Delta\rightarrow 0$ and
our analysis is not applicable.  This is a consequence of the van Hove
singularities at $(\pm \pi,0)$, $(0,\pm \pi)$ in that model.  Schulz
\cite{Schulz} and Dzyaloshinskii \cite{Dzyaloshinskii} applied a
complementary method to treat the tight binding spectrum: they showed that
the BCS and CDW susceptibilities are governed by electrons in the corners of
the Fermi surface, which led them to replace the Fermi surface by 4 points
and they treated this simplified model by renormalization group methods.
Similarly as in Refs.\cite{Schulz,Dzyaloshinskii}, Houghton and Marston
\cite{HM} model a tight binding spectrum close to half filling by 4 points
which, however, represent the branches of the spectrum and van Hove
singularities are neglected. In this respect, their model is similar to
ours.  For spinless fermions with repulsion, Houghton and Marston find, in
absence of large momentum scatterings, no instability towards CDW state
formation in agreement with our result. The model Eq.(1) bears some
resemblance also to the problem studied in Ref.\cite{Guinea}, where
conclusions similar to ours were reached.

In conclusion, we studied a two-dimensional model (generalization to higher
D is straightforward), where a Luttinger liquid is realized. The model
exhibits an analogue of spin-charge separation; the role of spin is played
by the momentum transverse to the direction of the quasi-one-dimensional
motion and the electron Green's function is expected to have new features
compared to $D=1$. It will be interesting to study the influence of
curvature of the Fermi surface, electron spin and large momentum scatterings
on the stability of Luttinger liquids.

%
%
%
%Acknowledgement
%
%
%
We would like to thank Prof. T. M. Rice for continuous support and
illuminating discussions. We thank also S. Gopalan, D. Khveshchenko and H.
Tsunetsugu for helpful conversations.

%
%
%
%REFERENCES
%
%
%

%
%
%
%Figures
%
%
%

\begin{figure}
\caption{Fermi surface of the model Eq.(1). The branches are
labelled 1-4. $\Lambda$ is the length of the Fermi surface in one branch.
${\bf Q}_1$ and ${\bf Q}_2$ are nesting vectors of the Fermi surface.}
\end{figure}

\end{document}